\documentclass[aps,prb,twocolumn,showpacs,superscriptaddress]{revtex4-1}
\usepackage{graphicx}

\begin{document}

\title{Spin ordering and electronic texture in the bilayer iridate Sr$_3$Ir$_2$O$_7$}

\author{Chetan Dhital}
\thanks{These authors contributed equally to this work.}
\affiliation{Department of Physics, Boston College, Chestnut Hill, Massachusetts 02467, USA}
\author{Sovit Khadka}
\thanks{These authors contributed equally to this work.}
\affiliation{Department of Physics, Boston College, Chestnut Hill, Massachusetts 02467, USA}
\author{Z. Yamani}
\affiliation{ Chalk River Laboratories, Canadian Neutron Beam Centre, National Research Council, Chalk River, Ontario, Canada K0J 1P0} 
\author{Clarina de la Cruz}
\affiliation{ Neutron Scattering Science Division, Oak Ridge National Laboratory, Oak Ridge, Tennessee 37831-6393, USA}
\author{T. C. Hogan}
\affiliation{Department of Physics, Boston College, Chestnut Hill, Massachusetts 02467, USA}
\author{S. M. Disseler}
\affiliation{Department of Physics, Boston College, Chestnut Hill, Massachusetts 02467, USA}
\author{Mani Pokharel}
\affiliation{Department of Physics, Boston College, Chestnut Hill, Massachusetts 02467, USA}
\author{K. C. Lukas}
\affiliation{Department of Physics, Boston College, Chestnut Hill, Massachusetts 02467, USA}
\author{Wei Tian}
\affiliation{ Neutron Scattering Science Division, Oak Ridge National Laboratory, Oak Ridge, Tennessee 37831-6393, USA}
\author{C. P. Opeil}
\affiliation{Department of Physics, Boston College, Chestnut Hill, Massachusetts 02467, USA}
\author{Ziqiang Wang}
\affiliation{Department of Physics, Boston College, Chestnut Hill, Massachusetts 02467, USA}
\author{Stephen D. Wilson}
\email{stephen.wilson@bc.edu}
\affiliation{Department of Physics, Boston College, Chestnut Hill, Massachusetts 02467, USA}

\begin{abstract}
Through a neutron scattering, charge transport, and magnetization study, the correlated ground state in the bilayer iridium oxide Sr$_3$Ir$_2$O$_7$ is explored. Our combined results resolve scattering consistent with a high temperature magnetic phase that persists above $600$ K, reorients at the previously defined $T_{AF}=280$ K, and coexists with an electronic ground state whose phase behavior suggests the formation of a fluctuating charge or orbital phase that freezes below $T^{*}\approx70$ K.  Our study provides a window into the emergence of multiple electronic order parameters near the boundary of the metal to insulator phase transition of the 5d $J_{eff}=1/2$ Mott phase.     
\end{abstract}


\maketitle

There has recently been considerable interest in studying the phase behavior of correlated 5d-electron transition metal oxides due to the potential of realizing fundamentally new electronic phenomena where electron hopping, spin-orbit coupling, and Coulomb interaction energy scales are almost equivalent \cite{yang, pesin, wang}.  Of particular focus have been members of the iridium oxide Ruddelsden-Popper (RP) series Sr$_{n+1}$Ir$_n$O$_{3n+1}$ where an experimental picture of a novel spin-orbit induced $J_{eff}=1/2$ Mott insulating state has been proposed \cite{kim214, kimscience}. Upon increasing the dimensionality of the iridate RP series to higher $n$, optical \cite{moon} and transport measurements \cite{longo, cao327} have shown that the effective bandwidth increases and the system transitions from a quasi-two dimensional insulating state to a metallic phase in the three dimensional limit.  

Specifically, the reported optical gap in the $n = 2$ member Sr$_3$Ir$_2$O$_7$ (Sr-327) shifts considerably downward relative to the $n=1$ Sr$_2$IrO$_4$ system into what should be a weakly insulating phase \cite{moon}---demonstrating that Sr-327 occupies a unique position in the iridate RP phase diagram near the boundary of the metal to insulator phase transition in the RP series.  Given this framework, Sr$_3$Ir$_2$O$_7$ exhibits a number of anomalous features in its magnetic properties: Bulk magnetization measurements of Sr-327 reveal a rich behavior possessing three distinct energy scales \cite{cao327,mcmorrow}, and recent $\mu$sR measurements have revealed the presence of highly disordered local spin behavior \cite{franke}; both supporting the notion of multiple coexisting or competing magnetic phases.  However, the details of how spin order evolves in this material and interfaces with the energy scales identified in both transport and bulk susceptibility measurements remains largely unexplored.

In this article, we utilize neutron scattering, bulk magnetization, and transport techniques to explore the phase behavior in Sr$_3$Ir$_2$O$_7$ (Sr-327).  At high temperatures, a previously unreported phase appears with $T_{onset}>600$ K followed by a second magnetic transition at $T_{AF}=280$ K. Scattering from this high temperature phase is consistent with a magnetic origin, provides an explanation for the absence of Curie-Weiss paramagnetism in this material above $280$ K \cite{nagai}, and also suggests an origin for the recently reported anomalous $93$ meV magnon gap \cite{kimmagnongap}. At low temperatures, the spin order is decoupled within resolution from a second upturn in the bulk spin susceptibility at $T_{O}=220$ K suggestive of the formation of an electronic glass that freezes below T$^{*}\approx70$ K.  Below this freezing energy scale, charge transport demonstrates a localized ground state that can be biased into a regime of field enhanced conductivity (FEC) consistent with collective transport above a threshold electric field.  Our combined results demonstrate the coexistence of spin order with an unconventional, electronically textured, phase in an inhomogeneous ground state near the boundary but on the insulating side of the $J_{eff}=1/2$ Mott transition.

\begin{figure}
\includegraphics[scale=.3]{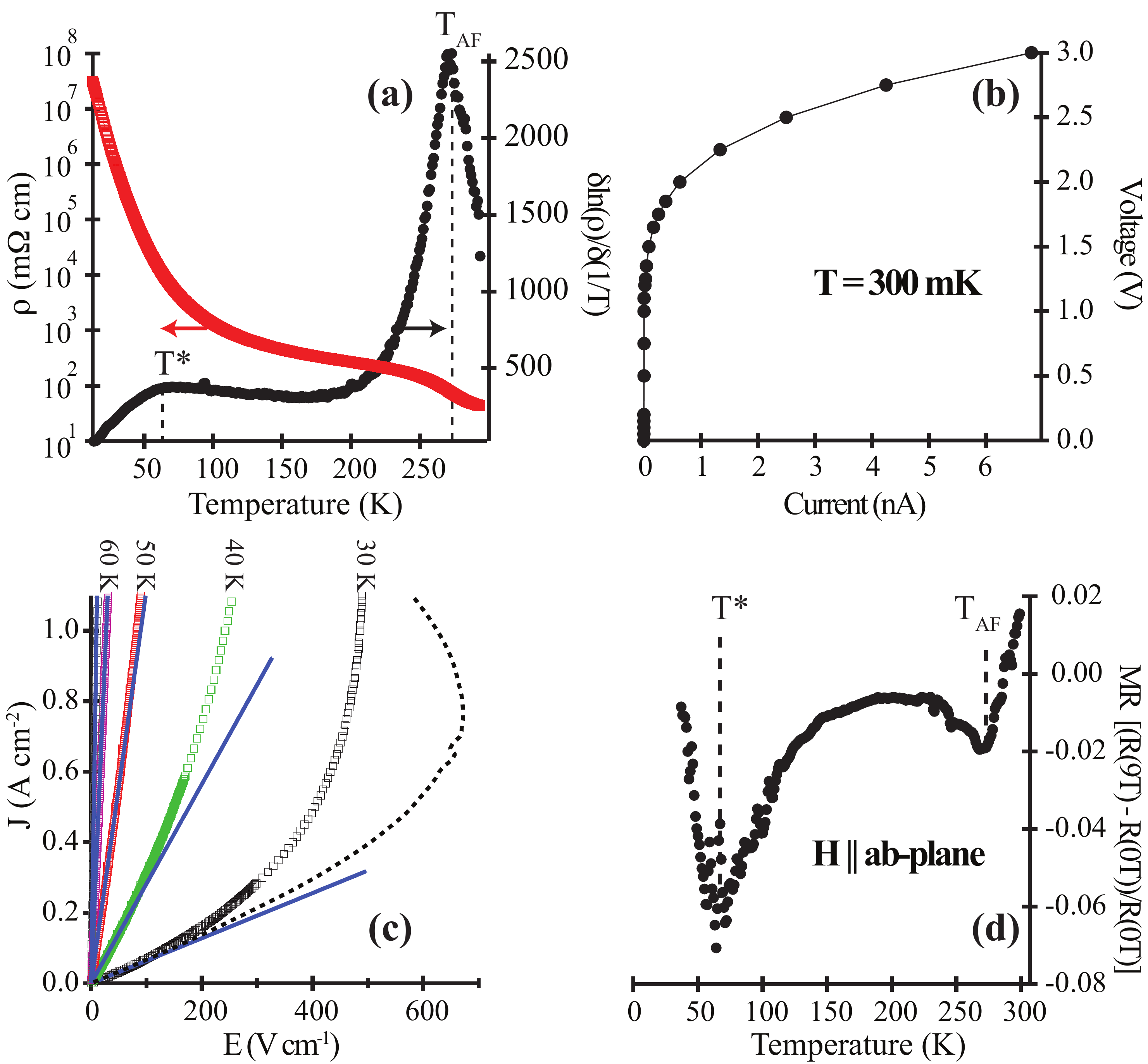}
\caption{(a) Temperature dependence of the $ab$-plane resistivity for Sr-327.  Also plotted is the $\frac{\partial ln\rho}{\partial (1/T)}$ versus $T$ showing two peaks at $T_{AF}$ and $T^{*}$   (b) IV curve of $ab$-plane transport at 300 mK showing voltage biasing into a FEC regime.  (c) Current driven, pulsed, IV measurements as a function of temperature.  Solid lines show linear fits to the Ohmic regime at each temperature.  Dashed line is joule heating model at 30K described in the text. (d) MR ratio as described in the text plotted as a function of temperature showing two well defined minima at the $T*$ and $T_{AF}$ transitions.}   
\end{figure}

Single crystals of Sr$_3$Ir$_2$O$_7$ (Sr-327) were grown via flux techniques similar to earlier reports \cite{subramarian, supplemental}.  The resulting Sr:Ir ratio was confirmed to be 3:2 via EDS measurements, and a number of Sr-327 crystals were also ground into a powder and checked via X-ray diffraction in a Bruker D2 Phaser system. No coexisting Sr$_2$IrO$_4$ phase was observed and the resulting pattern was refined to the originally reported I4/mmm structure---we note however that, due to the small scattering signal from oxygen, we are unable to distinguish between this and the various reported orthorhombic symmetries \cite{subramarian, cao327, nagai}. For the remainder of this paper, we will index the unit cell using the pseudo-tetragonal unit cell with with $a=b=5.50\AA$ $c=20.86\AA$. 

Neutron measurements were performed on the HB-1A triple-axis spectrometer at the High Flux Isotope Reactor (HFIR) at Oak Ridge National Laboratory and on the C5 spectrometer at the Canadian Neutron Beam Centre at Chalk River Laboratories.  Experiments on C5 were performed with a vertically focusing pyrolitic graphite (PG-002) monochromator and analyzer, an E$_f = 14.5$ meV, two PG filters after the sample, and collimations of 33$^{\prime}$-48$^{\prime}$-51$^{\prime}$-144$^{\prime}$ before the monochromator, sample, analyzer, and detector respectively.  On HB-1A, a double bounce PG monochromator was utilized with fixed E$_i=14.7$ meV, two PG filters before the sample, and collimations of 48$^{\prime}$-48$^{\prime}$-40$^{\prime}$-68$^{\prime}$.  Magnetization measurements were performed on a Quantum Design MPMS-XL system and resistivity data was collected in a series of four-wire setups: (1) zero field resistance from 300 K to 12 K was collected with a Keithley 2182A voltmeter, (2) data from 12 K to 0.3 K was collected in a $^{3}$He absorption refrigerator with an Keithley Model 617 electrometer, and (3) magnetoresistance data was collected in a 9T Quantum Design PPMS.

Looking first at the results of our $ab$-plane transport measurements under low (1 $\mu$A) current, Fig. 1 (a) shows the zero field resistivity as a function of temperature.  The sample's resistivity increases from several m$\Omega$-cm at room temperature to beyond 10 M$\Omega$-cm below 20 K and begins to show saturation behavior below 2 K \cite{supplemental}. There is no substantial interval of constant activation energy as illustrated by the overplot of $\frac{\partial ln\rho}{\partial (1/T)}$ versus $T$ in this same panel.  Instead, $\frac{\partial ln\rho}{\partial (1/T)}$ shows two peaks suggestive of two phase transitions coupling to charge carriers: the first near the known magnetic phase transition at $T_{AF}=280$ K \cite{cao327} and the second indicating a lower temperature phase formation at $T^{*}\approx 70$ K.       
        
\begin{figure}
\includegraphics[scale=.3]{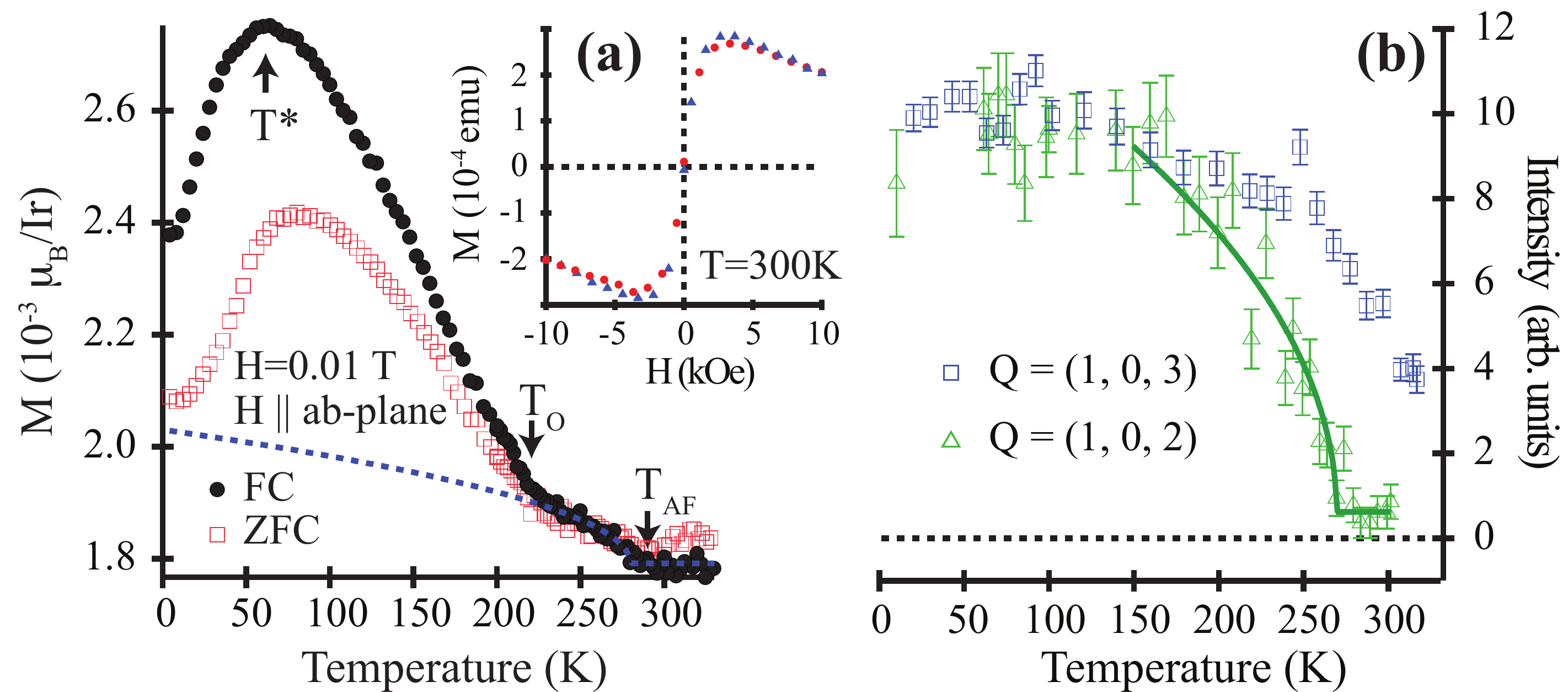}
\caption{(a) DC-magnetization data for Sr-327 with $H=0.01$ T aligned parallel to the ab-plane for both FC (solid symbols) and ZFC temperature sweeps (open symbols).  Dashed line shows mean-field order parameter fit to the net moment from the $280$ K transition.  Inset shows M versus H sweep at 300 K. (b) Temperature dependence of the peak intensities at (1,0,3) and (1,0,2) magnetic reflections. Solid line is a power law fit to the (1,0,2) order parameter.}
\end{figure}

In order to investigate the transport properties of this lower temperature, $T^{*}$ phase further, the charge transport was characterized via a voltage driven IV sweep at 300 mK shown in Fig. 1 (b). A pronounced nonlinearity appears, where with increasing field strength the system switches from a linear, Ohmic, regime with near zero conductance into a highly non-Ohmic FEC regime. To determine the temperature evolution of this FEC feature, a seperate sample was mounted and probed with $600$ $\mu$s current pulses to minimize heating effects (Fig. 1 (c)). While it is difficult to completely preclude all heating effects within the rise/sample time of the pulse, these pulsed measurements show that the nonlinear bend in the IV curve persists and eventually vanishes below resolution at $T\approx60$ K. 

A seperate (rough) check for discriminating the nonlinear conduction from simple joule heating can be performed by looking at the $30$ K data in Fig. 1 (c).  The Ohmic-regime $R$($30$ K)=$42$ k$\Omega$ and the maximum pulsed current ($2$ mA) during the 600 $\mu$s pulse delivers a maximum $\Delta Q=10.1\times10^{-5}$ J.  While no low temperature heat capacity data have been published for Sr-327, as a lower estimate, the heat capacity of Sr$_2$IrO$_4$ at 30 K can be used ($\approx 14$ J/K) \cite{kini} giving a maximum $\Delta T=5.5$ K (for a $1.32\times10^{-6}$ mole sample).  In carrying out a similar analysis for each current value pulsed at $30$ K and assuming \textit{perfect} thermal isolation, the measured Ohmic $R(T)$ can be used to determine the lowest fields possible due to pure joule heating as a function of the pulsed current density.  This limiting case is plotted as a dashed line in Fig. 1 (c)---demonstrating that the nonlinear feature at $30$ K is intrinsic \cite{note1}.       

\begin{figure}
\includegraphics[scale=.3]{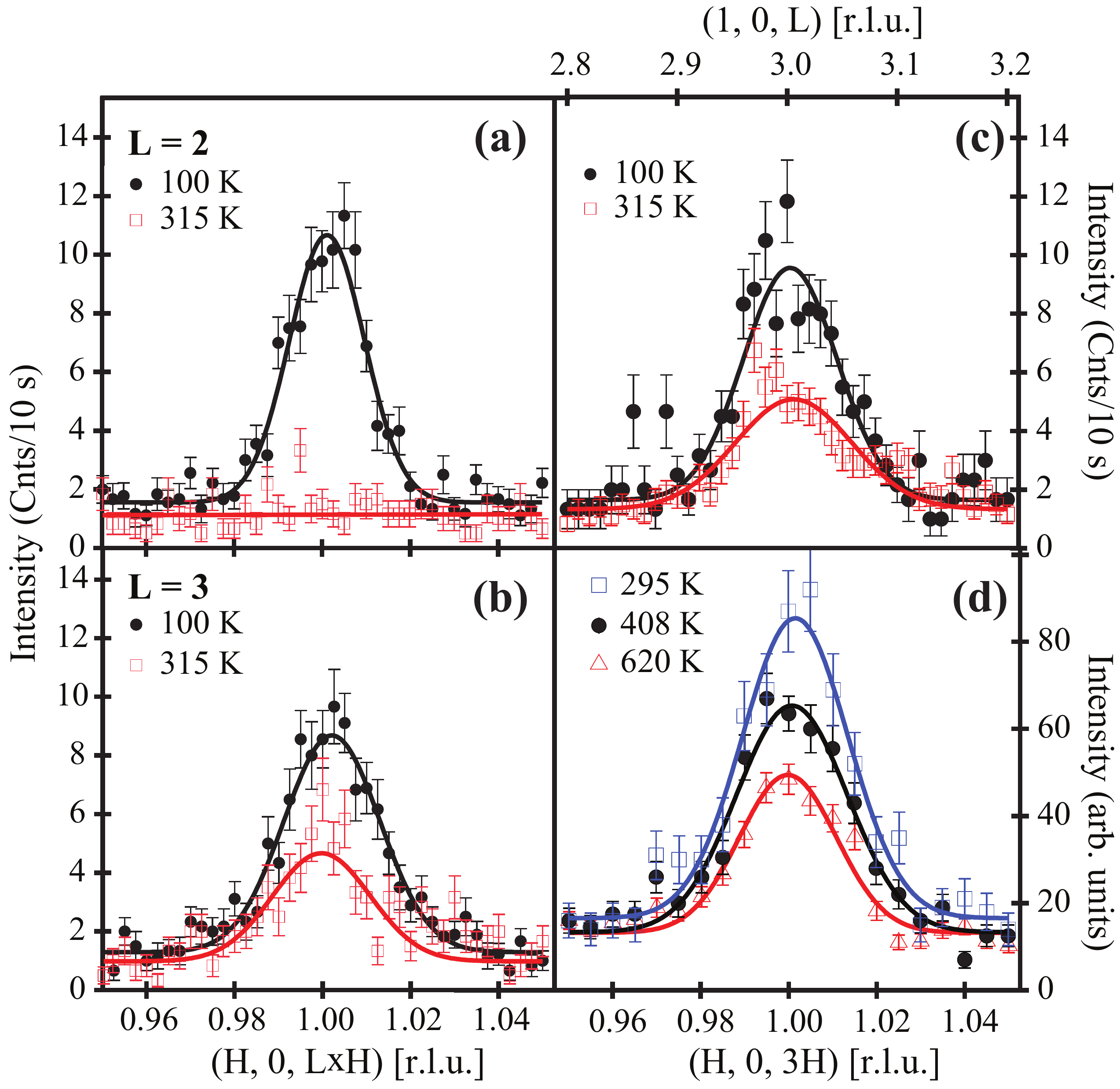}
\caption{Radial Q-scans at 100 K and 315 K through the (a) Q=(1,0,2) (b) and Q=(1,0,3) reflections.  Solid lines are Gaussian fits to the data. (c) L-scans across the (1, 0, 3) peak position showing 3D AF at 100 K and 315 K. (d) Q-scans showing the temperature dependence of the (1,0,3) peak.}
\end{figure}  

In looking at the magnetoresistance of the same sample plotted in the Fig. 1 (d), the MR=(R(9T)-R(0T))/R(0T) ratio is negative and shows two minima at $T^{*}\approx70$ K and $T_{AF}=280$ K.  The lower minimum appears approximately at the temperature where the onset of FEC emerges and coincides with the low-T peak in $\frac{\partial ln\rho}{\partial 1/T}$.  The origin of the negative magnetoresistance is likely the removal of spin disorder scattering due to biased magnetic domain populations which will be discussed later; and the inflection below $T^{*}$ supports the idea of a field coupled order parameter freezing below $70$ K.  The suppression of enhanced fluctuations originating from an additional electronic instability however may also account for the overall negative MR. 

Magnetization data shown in Fig. 2 (a) supports the idea of a bulk phase transition below $70$ K where a downturn in the dc-susceptibility originally reported by Cao et al. \cite{cao327} begins, suggestive of a glassy freezing process.  Consistent with earlier reports \cite{cao327, mcmorrow}, three energy scales are apparent in the field cooled magnetization data:  a canted AF phase transition at $T_{AF}=280$ K, a sharp upturn at $T_{O}=220$ K, and an eventual decrease in susceptibility below $T^{*}=70$ K.  Both field cooled (FC) and zero field cooled (ZFC) data show similar downturns near $T^{*}$ and an irreversibility temperature near $T_{O}$. At $300$ K however, field sweeps plotted in the inset of Fig. 2(a) reveal a rapid saturation of the spin response suggesting the persistence of magnetic correlations above $T_{AF}$.   

In order to further investigate the spin order, neutron diffraction measurements were performed on a $7$ mg single crystal Sr-327 sample with the results plotted in Figs. 3 and 4.  $[H,0,L]$, $[H,K,0]$, and $[H,H,L]$ zones were explored and magnetic reflections were observed only at the $(1, 0, L)$ positions for $L=1,2,3,4,5$.  The correlated order is three dimensional with $\xi_{L}=\sqrt{2ln(2)}\times1/w=147\pm10\AA$  where $w[\AA^{-1}]$ is the peak's Gaussian width (Fig. 3 (c)). The appearance of both $L=even$ and $L=odd$ reflections in a simple collinear picture of the spin structure is therefore consistent with recent X-ray results resolving the presence of two magnetic domains \cite{mcmorrow}---attributable to in-plane structural twinning in an orthorhombic symmetry.              

Looking at the order parameters for both the $L=3$ and $L=2$ reflections in Fig. 2 (b), the magnetic intensities show that the $L=2$ peak disappears at $T_{AF}$ while substantial intensity remains at $280$ K in the $L=3$ reflection. $\textbf{Q}$-scans plotted in Fig. 3(b) demonstrate this more explicitly.  The peak remaining above $280$ K is long-range ordered with a minimum correlation length of $93\pm18\AA$---comparable to the correlation length observed at $10$ K ($97\pm5\AA$). Due to the rather coarse collimations used, both these values and those of all magnetic Bragg reflections are resolution limited.  At $300$ K peaks remain at the $(1,0,L)$ $L=1,3,4$ positions, all forbidden in the reported structural space groups to date.  This same crystal was then loaded into a furnace and measured at higher temperatures, where, upon warming, the remnant peaks continue to decrease in intensity as illustrated in Fig. 3(d); however they notably remain present beyond $600$ K.  The continued temperature dependence of these peaks above $300$ K and the absence of peaks at higher order $L$ and $H$ strongly imply that this remnant scattering is magnetic and that an additional magnetic phase persists beyond $280$ K.

In order to verify this in a second sample, a $2$ mg crystal from a seperate batch was explored on the C5 spectrometer with the results plotted in Fig. 4 (a).  Again, a clear temperature dependence above 300 K was observed with the remnant $(1,0,3)$ peak vanishing within the error of the measurement by $450$ K.  The earlier disappearance of this high temperature AF peak is likely due to the poorer statistics in the measurement of this second sample; however variable oxygen stoichiometry between samples may also play a role in diminishing the effective transition temperature. 

\begin{figure}
\includegraphics[scale=.3]{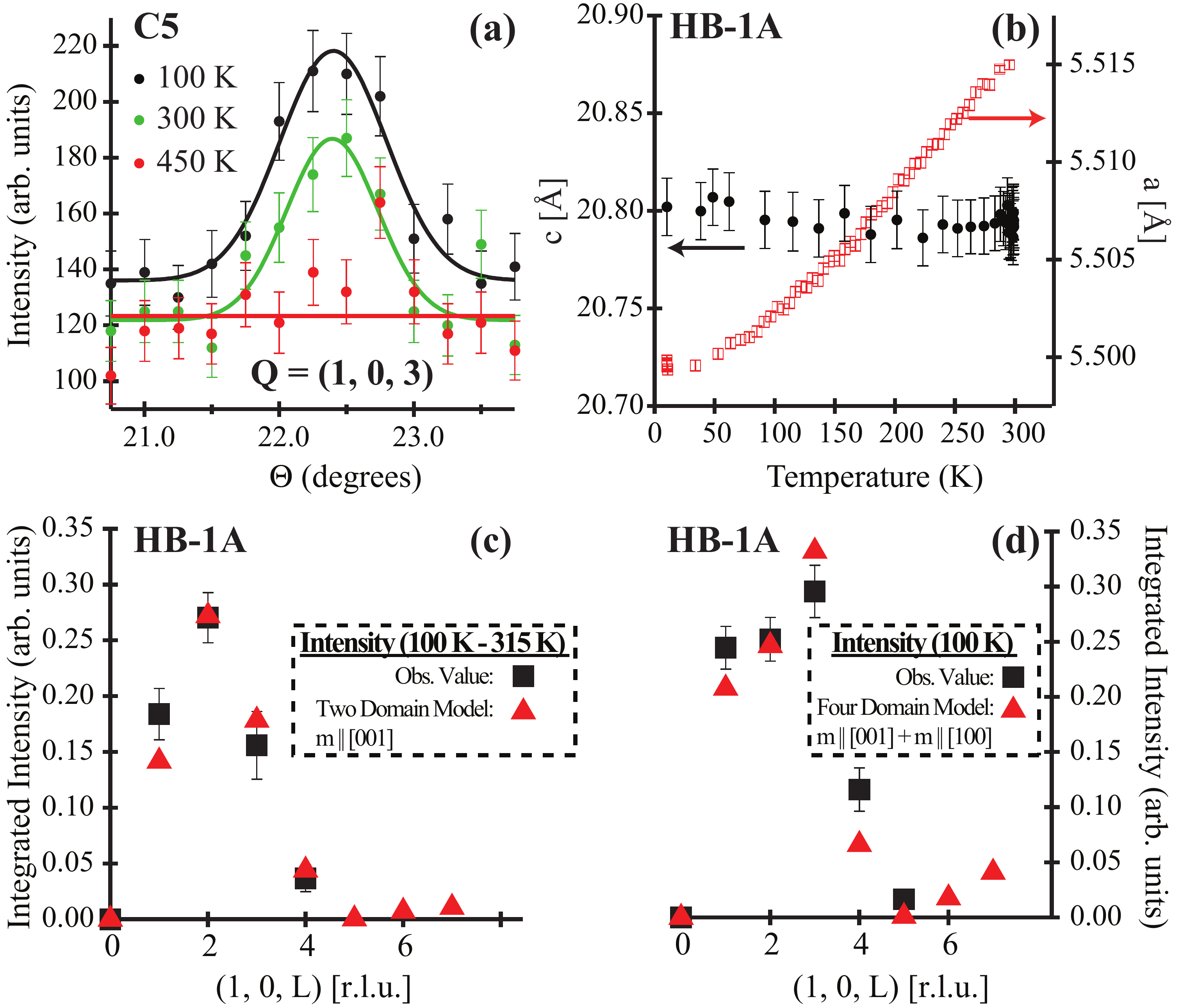}
\caption{(a) Rocking scans on a seperate crystal showing the temperature dependence of the (1,0,3) peak above 300 K.  (b) Temperature dependence of the a- and c- axis lattice parameters measured at the (2,0,0) and (0,0,4) reflections.  Integrated intensities plotted as (c) 100 K with 315 K data subtracted and (d) the total scattering data at 100 K.  Data is compared with two simple collinear spin models described in the text.}
\end{figure}

Due to the presence of two magnetic domains \cite{mcmorrow} and the rapid attenuation due to the Ir magnetic form factor, it is difficult to uniquely determine a model of the spin structure in both the high and low temperature magnetic phases. If we assume that the scattering seen at $315$ K is a seperate, saturated, order parameter, then the additional intensity due to the $280$ K transition is plotted in Fig. 4(c).  The rapid disappearance of magnetic peaks for $L>5$ suggests a sizable component of the moment directed along the c-axis, and the best symmetry bound two-domain model matching the data is a G-type, arrangement of AF-coupled bilayers with moments directed along the $(0,0,1)$-axis consistent with a recent X-ray report \cite{kim327, supplemental}. The ordered moment using this model is $\mu=0.52\pm 0.08$ $\mu_{B}$.

Looking instead at the total scattering observed at $100$ K in Fig. 4 (d), no simple collinear model captures all of the major reflections well. Nevertheless, if we again use a twinned G-type spin structure, a model comprised of four magnetic domains with two different moment orientations can be constructed. If the two new twin domains added to the previous model have moments directed along the $(1,0,0)$-axis, this four domain model roughly fits the data \cite{supplemental}.  This added domain would comprise the high temperature phase in a two domain picture; however future polarized measurements are required to differentiate between this multidomain picture, a potential noncollinear spin structure with an accompanying spin reorientation at $280$ K, and to confirm the magnetic nature of the high temperature phase.

Our combined data demonstrate the presence of canted 3D antiferromagnetic domains whose phase evolution is decoupled within resolution from the fluctuation/freezing behavior at $T^{*}$ and $T_{O}$ (Fig. 2 (b)); precluding any additional major spin reorientations at these temperatures.  This suggests that there remain additional moments weakly coupling \cite{cao327,mcmorrow} to fluctuations below $T_{O}$ and eventually freezing below $T^{*}$. Our measurements in their entirety therefore suggest a picture of three distinct order parameters driving the phase behavior of Sr-327:  (1) a high temperature phase (of likely magnetic origin) with $T_{onset}>620$ K, (2) a canted AF magnetic transition at $280$ K followed by (3) the freezing of the $T^{*}$ phase into an electronically textured ground state.  
    
The $T^{*}$ transition is nominally suggestive of a charge density wave or collective transport mechanism which becomes depinned above a threshold field---leading to an avalanche process in the carrier number. The structural lattice parameters (Fig. 4 (b)) however evolve smoothly as the system is cooled from $315$ K to $10$ K and, to date, no structural distortion associated with a conventional CDW formation has been observed below $300$ K \cite{nagai, matsuhata}. Although, to the best of our knowledge, high temperature structural studies have yet to be reported.  An alternative scenario of exchange coupled metallic islands condensing below $T^{*}$ with a substantial Coloumb barrier for tunneling may also address the transport mechanism below $T^{*}$ \cite{arovas,jose}.  Similar non-Ohmic behavior has also been reported in other correlated iridates \cite{cao214, cao113} suggesting an electronic inhomogeneity intrinsic to these 5d-correlated materials.    

Curiously, X-ray measurements on a Sr-327 sample with a qualitatively similar bulk spin susceptibility have reported the onset of AF order at $T_{O}$ \cite{mcmorrow}. This RXS study speculated about the presence of short-rage order setting in at $T_{AF}$ and diverging at $T_O$ as the reason for the discrepancy \cite{mcmorrow}; however our measurements reveal no appreciable change in the correlation length upon cooling through $T_O$.  Given that more recent RXS measurements show the onset of magnetism at the expected $T_{AF}=285$ K\cite{kim327}, variation in sample quality is likely the cause for the variance reported between these two RXS studies.

To summarize, our studies have illustrated a complex electronic ground state in the Sr$_3$Ir$_2$O$_7$ system with multiple electronic order parameters. Our observation of scattering consistent with an AF phase extending beyond $600$ K is supported by the absence of Curie-Weiss behavior above the previously identified $T_{AF}$ \cite{nagai} and also by the rapid field-induced saturation of the  magnetization at $300$ K.  The system then transitions through a magnetic transition at $T_{AF}=280$ K, and exhibits multiple magnetic domains or alternatively noncollinear spin order in its ground state.  The spin order appears decoupled from two additional energy scales appearing in transport and bulk susceptibility measurements, suggesting a fluctuating charge/orbital state that freezes into an inhomogeneous electronic ground state where tunneling/sliding effects manifest under increasing electric field strength.

\acknowledgments{
SDW acknowledges helpful discussions with Ying Ran and Stefano Boseggia, and Michael Graf for use of a $^3$He refrigerator.  The work at BC was supported by NSF Award DMR-1056625 and DOE DE-SC0002554.  Part of this work was performed at ORNL's HFIR, sponsored by the Scientific User Facilities Division, Office of Basic Energy Sciences, U.S. DOE.}


\begin{thebibliography}{}
\bibitem{note1} Both higher field pulsed measurements and DC measurements show pronounced regions of negative differential resistance; however for the purpose of this paper we limit our discussion to only measurements where the effect of joule heating has been minimized.
\end{thebibliography}

\end{document}